
\input phyzzx
\def\co{cohomology\ }
\def\G{${G\over G}\ $}
\def\GT{${G\over G}$ TQFT\ }
\def\F{{\cal F}(J,I)\ }
\def\c#1#2{\chi_{#1}^{#2}}
\def\p#1#2{\phi_{#1}^{#2}}
\def\b#1{\beta_{#1}}
\def\g#1{\gamma_{#1}}
\def\r#1#2{\rho_{#1}^{#2}}
\def\jt#1#2{{J^{(tot)}}_{#1}^{#2}}
\def\y{|phys>}
\def\t{\tilde}
 \def\pa{\partial}
\def \f{f^a_{bc}}
\def \fcr{\f \c b {} \r c {} }
\def \jg{J^{(gh)}}
\def \jb{J^{(BRST)}}
\def \p{\partial}

\def\cmp#1{{\it Comm. Math. Phys.} {\bf #1}}

\def\pl#1{{\it Phys. Lett.} {\bf #1B}}
\def\prl#1{{\it Phys. Rev. Lett.} {\bf #1}}
\def\prd#1{{\it Phys. Rev.} {\bf D#1}}

\def\np#1{{\it Nucl. Phys.} {\bf B#1}}

\def\jmath#1{{\it J. Math. Phys.} {\bf #1}}
\def\mpl#1{{\it Mod. Phys. Lett.}{\bf A#1}}
\def\jmp#1{{\it J. Mod. Phys.}{\bf A#1}}

\REF\Wym{E. Witten, \cmp {117} (1988) 353.}
\REF\BKDSGM{D. Gross and A. A. Migdal \prl {64} (1990) 127;\break
M. Douglas and S. Shenker \np {335} (1990) 635;\break E. Brezin and V. A.
Kazakov, \pl {236} (1990) 144}
\REF\DDK{F. David \mpl {3}  (1988) 1651;\break J. Distler and H. Kawai \np
{321}
(1989) 509.}
 \REF\Wrr{E. Witten, \np {340} (1990) 281.}
\REF\SY{M. Spigelglas and S. Yankielowicz `` \G Topological Field Theories by
Coseting $G_k$ TAUP-1934
;``Fusion Rules As Amplitudes in $G/G$ Theories,'' Technion PH- 35-90}
\REF\Wgg{E. Witten, ``On Holomorphic Factorization of WZW and Coset Models"
IASSNS-91-25.  }
 \REF\GK{
  K. Gawedzki and A. Kupianen , \pl {215} (1988) 119, \np
     {320} (1989)649.}
\REF\KS{D. Karabali and H. J. Schnitzer, \np {329} (1990) 625.}
\REF\MS{D. Montano, J. Sonnenschein ,  \np {324} (1989) 348,
J. Sonnenschein \prd {42} (1990) 2080.}
\REF\Wcs{E. Witten \cmp {121} (1989) 351.}
\REF\BerK{M. Bershadsky and I. Klebanov \np {360} (1991) 559.}
 \REF\LZ{B. Lian and G. Zukerman \pl   {254}  (1991) 417.}
\REF\BMP{P. Bouwknegt, J. McCarthy and  K. Pilch Cern Preprint
TH-6162/91.}
\REF\Wgr{E. Witten ``Ground Ring of Two Dimensional String Theory" IASSNS-HEP
91/51}
\REF\Kleb{  I. R. Klebanov,  Princeton preprint PUPT-1302.}
\REF\KMS{D. Kutazov, E. Martinec and N. Seiberg  Princeton preprint PUPT-1293.}
\REF\GKN{D. J. Gross, I. R. Klebanov and M. J. Newman \np {350} (1991) 621.}
\REF\BO{M. Bershadsky and H. Ooguri \cmp {126}  (1989) 49.}
\REF\KK{V. G. Kac and  D. A. Kazhdan   {\it Adv. Math } {\bf  34}  (1979) 79.}
 \REF\MFF{F. G. Malinkov, B.L.  Feigin and D. B.  Fuks{\it  Funkt. Anal.
Prilozhen} {\bf 20 No. 2} (1987) 25. } \REF\BT{ R. Bott L. W. Tu
``Differential
Forms  in Algebric Topology", Springer-Verlag  NY  1982.}
\REF\BF  {Bernard   and G, Felder \cmp {}  (1991)  145.}
\REF\Bersh{This case  was also analyzed by
M.Bershadsky [private communication]}.\REF\WN{A. Zamolodchikov,
{\it Teor. Mat. Phys.} {\bf 65} (1985) 1205\break
V. Fateev and S. Lukyanov, \jmp {3} (1988) 507.}
\REF\Wgra{ See for example K. Schoutens, A. Servin and P. van Nieuwenhuizen,
``Properties of Covariant W Gravity" in proceeding of the June 1990 Trieste
Conference on ``Topological Methods in Field Theory", and reference therein.}
\REF\Wak{M. Wakimoto,   \cmp   {104}  (1989) 605.}
\REF\MO{N. Marcus and Y. Oz "Discrete States of 2D String Theory in Polyakov's
Light-Cone
Gauge". Tau preprint TAUP-1962-92.}
 \REF\Pol{A. M. Polyakov, \mpl  {2}  (1987) 893.}
\REF\KPZ{V. G. Knizhnik, A. M. Polyakov and A. B. Zamolodchikov   \mpl  {3}
(1988) 819.}
\REF\WD{R. Dijkgraaf, E. Witten, \np {342} (1990) 486.}
\REF\dvv{R. Dijkgraaf, E. Verlinde, and H. Verlinde,
\np {352} (1991) 59;
``Notes On Topological String THeory And 2D Quantum Gravity,'' Princeton
preprint PUPT-1217 (1990).}
\REF\PV{A. Polyakov and P. B. Wigmann \pl {131} (1983)  121.}
\REF\FGPP{P. Furlan, A. Ch. Ganchev, R. Paunov and V. B. Petkova Cern preprint
CERN-TH-6289/91.}
 \REF\Gepner{D. Gepner \cmp {141} (1990) 381.}
\REF\NW{D. Nemeschansky and N. P. Warner ``Topological Matter, Integrable
Models
 and Fusion Rings"  USC-91/031.}
\REF\Spigel{``Setting Fusion Rules in Landau-Ginzburg Space" Technion preprint
PH-8-91 {\it Phys. Lett} to appear.}
\REF\NY{D. Nemeschansky and S. Yankielowicz ``N=2 W algebras, Kasama-Suzuki
Models and Drinfeld-Sokolov Reduction"
   USC-007-91.}
\REF\FeFr{B. Feigin and E. Frenkel \pl {246} (1990) 75.}
\REF\RCW{A. Rocha-Caridi and N. R. Wallach, Trans. Am.
Math. Soc. 277 No.1 (1983)133. }
\REF\FF{ B. L. Feigin and P. B. Fuks  ``Representations of the Virasoro
Algebra" Seminar on Supermanifolds  {\bf 5} ed. D. Leites.}
\REF\BS{ M. Bauer and N. Sochen, \pl {275} (1992)82
M. Bauer and N. Sochen, ``Fusion and Singular Vectors in $A^{(1)}_1$ highest
weight cyclic modules'' hepth/9201079 }.

\rightline{TAUP- 1961-92}
\date{April 1992}
\titlepage
\vskip 1cm
\title{Physical States in \G Models and  2d Gravity}
\author {O. Aharony, O. Ganor , J. Sonnenschein and S. Yankielowicz
\footnote{\dagger}{Work supported in part by the US-Israel Binational Science
Foundation and the Israel Academy of Sciences.}}
\address{ School of Physics and Astronomy\break
Beverly and Raymond Sackler \break
Department of Exact Sciences\break
Ramat Aviv Tel-Aviv, 69987, Israel}
\author {N. Sochen}
\address{Service de Physique Theorique\break
Cen Saclay\break
91191 Gif-sur Yvette Cedex, France}
\abstract{ An analysis of the BRST cohomology of the \G
topological models is performed for the case of $A_1^{(1)}$.
Invoking a special free field parametrization of the  various
currents, the cohomology on the corresponding Fock space is
extracted.
We employ the singular vector structure and fusion rules to translate
the latter into  the cohomology on the space of  irreducible representations.
Using the physical states we calculate the characters and partition
function, and verify the index interpretation.
We twist  the energy-momentum tensor to establish an intriguing
correspondence between the ${SL(2)\over SL(2)}$ model with level
$k={p\over q}-2$ and $(p,q)$ models coupled to gravity.}
\overfullrule=0pt

\section{ Introduction}
Topological Quantum field thories\refmark\Wym have been a center of much
interest
during recent years. At the same time a great emphasis was devoted to the
research of non critical string theory for $c\leq  1$. This topic was explored
via matrix models\refmark{\BKDSGM} as well as continuum Liouville
theory.\refmark{\DDK} In several cases a strong relationship between  some
topological models and the string models was revealed. For instance recursion
relations for amplitudes in the one matrix model
were found to be identical to those derived in the topological theory of
gravity.\refmark\Wrr  The exploration of connections between another  class of
interesting  topological models, the so called \G  models, and  theories of
$c\leq 1$ matter  coupled to 2d gravity is the subject of the present work.

The \G topological theories\refmark{\SY,\Wgg} are   theories based on $G$
current algebra.
 Formally they are   obtained by gauging the $G$ WZW
model. Following the well known ${G\over H}$  construction\refmark{\GK,\KS}
 the \G models are
derived by gauging  the anomaly free group $G$ of the  $G_k$ WZW model.  This
recasts, using  a complexified $G^c$ algebra, the \G action into a sum of three
terms: a  $G$ WZW at level $k$, a $({G^c\over G})_{k+2C_G}$ WZW model which can
be viewed as a $G$ WZW model at level $-(k+2C_G)$ and a (1,0)  odd ghost system
in the adjoint representation of the group.
The \G\ models  have a significant overlap with other TQFT's. For one they are
related to the $2+1$ dimensional Chern-Simons theory\refmark\Wcs. In particular
the amplitudes, which are given in terms of $N_{ijk}$ fusion rule
coefficients,\refmark{\SY,\Wgg} coincide with the scalar product of wave
functions of the Chern-Simons theory. Other related TQFT  are the
topological flat connection models.\refmark\MS They share with the \G models
the
correspondence to the moduli space of flat gauge connections. More  interesting
is the relation with models of $c\leq1$ coupled to gravity. This is the main
topic of the present work.
 In ref. [\SY] the structure of the \G theories was investigated  and
some striking resemblance to the 2 dimensional gravity models was revealed.
One would expect  that the $G_k$ WZW model plays the role of the matter
system, while the $({G^c\over G})_{k+2C_G}$ WZW model  is affiliated
 with the Liouville degrees of freedom.
Both in the gravitational models\refmark\BerK as well as in the \G models the
full character
gets a contribution only from the   primary states of the matter ($G_k$)
sector whereas the matter and  Liouville ($G_{-(k+2C_G)}$) oscillators are
canceled out exactly by the ghost contribution. The character can be
interpreted
as an index which encodes the information about the BRST invariant states of
the
theory. This index is $Tr[(-)^Gq^N]$  and $Tr[(-)^Gq^N(e^{\pi
i\theta})^{J_3-\half }]$ for the gravity and   $A_1^{(1)}$
cases respectively, where $G$ stands here for the ghost number, $N$ is the
level
and $J_3$ is the charge, appropriately defined\refmark\SY.   The
 partition function is obtained by integrating over the moduli.
In the gravity case  this leaves only matter primary states while in the \G
model the partition function equals the number of conformal blocks. For the
$SU(2)$ case with an
integer level $k$ the result is $k+1$.
 The information from the index interpretation of the partition function
 was used in ref. [\SY]
to disentangle the spectrum. In view of possible cancelations,
an index cannot
a-priori  give full information about the spectrum.  A complete account of the
physical states can be achieved only by working out the cohomology of the
relevant BRST operator.
This type of analysis for  $c\leq 1$ matter coupled to gravity  was worked out
originally
 in ref. [\LZ] and latter in ref. [\BMP]. An extremely  rich spectrum
of
discrete states was revealed. They were found to correspond to previous matrix
model calculations\refmark\GKN In the case of $c=1$ a
`` ground ring"\refmark\Wgr
of discrete states  of zero ghost number and  dimension was exposed with an
underlying $W_\infty$ symmetry. The corresponding  Ward identities were
implemented in the computations of correlators.\refmark\Kleb  A different
method
for those calculations was  suggested in  ref. [\KMS] together with an
extension
to the $c<1$ models. The counterpart of these results in the context of the \G
models is still missing.
  One of the main tasks of the present work is to take the first step in this
direction. It  is devoted to a BRST analysis of the gauged  $SU(2)$ WZW
model
\footnote{\dagger}{Throughout the paper we shall freely move between
$SU(2)$ and $SL(2)$ cases depending on the values of the level $k$.}
   of level
$k$ defined on a sphere. We establish rigorously the result of ref. [\SY] that
for an integer $k$ there are $k+1$ zero ghost number primary states
corresponding to the matter primaries. On each of these primaries there is a
whole tower of descendant  states, one  at each ghost number. The outcome of
the  BRST analysis follows closely that of $c<1$ models coupled to 2d
gravity.\refmark{\LZ,\BMP} To be more precise $(p,q)$ minimal matter systems
coupled to gravity  correspond to $k={p\over q}-2$ level of the $SL(2)$ case.
This is related  and indeed has a lot in common with the observation that
$SL(2)$ WZW model is connected to  the  $(p,q)$ system via the Hamiltonian
reduction. \refmark\BO Moreover, it was proven\refmark {\BO} that the
irreducible representation of the Virasoro algebra can be extracted from the
irreducible representation of the $SL(2)$ current algebra by putting a
constraint. The quantum gravity coupled to $c_{p,q}$ matter was shown to be
equivalent to the constrained WZW model of level $-(k+4)$.

The paper is organized as follows. In section 2 the \G topological theory  is
briefly reviewed. This serves mainly to establish notation and put forward
important topics to what follows.
In section 3  a BRST analysis similar  to the one employed in ref.[\BMP] is
performed for
$G=A_1^{(1)}$. The main idea is to use a free field
parametrization for the currents of  the $k$ and $-(k+4)$ levels. In fact  it
turns out that to implement the procedure of ref.[\BMP] it is convenient to
invoke ``conjugate" parametrizations for the two sectors.
The physical states of the theory
correspond to  the cohomology on another space
${\cal
F}_k\otimes{\cal F}_{-(k+4)}\otimes {\cal F}_{ghost}$, where ${ \cal F}_k$
denotes the    irreducible representation  of  $A_1^{(1)}$ level $k$, ${\cal
F}_{-(k+4)}$ and ${\cal F}_{ghost}$ the Fock space of the $-(k+4)$ sector and
the ghosts respectively.
 In section 4 we discuss aspects of  the representation
theory of  Kac-Moody algebra $A^{(1)}_1$\refmark\MFF  for an
arbitrary complex level $k$. Special attention is devoted to singular vectors
associated with null states. The existence of these vectors in a given Verma
module indicates that it is reducible.\refmark\KK This information is
 crucial for
the determination of the BRST cohomology on the space of  irreducible
representations. There is a similarity between the representations of the
current algebra  and  those of the Virasoro algebra. The duality between
representations of the latter which are characterized by $|h,c>$ and
$|1-h,26-c>$ is shown\refmark\LZ to correspond to a duality between  $|J,k>$
and
$|-1-J,-(k+4)>$ in the $A^{(1)}_1$ algebra. The singular vectors are used in
the
appendix  to determine the fusion rules.
 The passage from the  Fock space BRST
cohomologies to an
irreducible representation is presented in section 5. It
 follows the work of Bernard and Felder\refmark\BF .
 In section 6 we demonstrate the
correspondence between $(p,q)$ minimal models coupled to gravity and the \G
model for $G=SL(2)$ and
 $k+2 ={p\over q}$. We  start with some ``numerological" indications
about the realtionship. Similar ``numerical" observations are also the basis
for the $SL(2)$ quantum Hamiltonian reduction approach\refmark{\BO}.
 We then show that we have the right number of
``primary" operators. Next  we suggest a map between the fields of the two
models after introducing a twisted energy momentum tensor. This  is
suplemented  by a computation of the partition function of the \G\
model and a proof that it coincides with that of the $(p,q)$ minimal model
provided a particular value of the moduli of flat $G$ gauge
connection is  picked.
An explicit construction of the operators which correspond to  physical states,
and their correlators is presented in section 7 for the simplest case of
$k=0$.\refmark\Bersh  This provides a demonstration of our
general considerations. We  write these operators down in two ``conjugate"
parametrizations, prove  that there is a one to one map between them and
determine   the non-vanishing correlation functions.
A summary and a set of
open questions is brought in section 8. We also present first results
indicating
that the generalizationto $G=SL(n)$ provides a model of $W_N$\refmark\WN matter
coupled to $W_N$\refmark\Wgra gravity. A derivation of fusion rules of the
rational level $k$ $A_1^{(1)}$ algebra using the singular vectors is presented
in
the appendix. 
\def \GW{$G-WZW$}
\def\jt#1#2{{J^{(tot)}}_{#1}^{#2}}
\def\y{|phys>}
\def\t{\tilde}
 \def\pa{\partial}
\def \f{f^a_{bc}}
\def \fcr{\f \c {} b \r  {} c}
\def \jg{J^{(gh)}}
\def \jb{J^{(BRST)}}
\def \p{\partial}
\section{ The \G Topological Field Theory}
The \GT is constructed by gauging the anomaly free diagonal $G$
group of the \GW model. The  classical action takes the form
$$ S_k(g,A^\mu) = S_{k}(g) -{k\over 2\pi}
\int_{\Sigma}d^2 z Tr(g^{-1}\p g \bar A_{\bar z} +  g\bar \p g^{-1} A_z -
\bar A_{\bar z}g^{-1}A_z g  + A_z\bar A_{\bar z} )\eqn\mishwzw$$
where $g\in G$ and $S_k(g)$ is the WZW action at level $k$. In case that
$\Sigma$ is topologically trivial the gauge  field can be parametrized as
follows
$A_z=ih^{-1}\p_z  h ,  \bar  A_{\bar z}=ih^*\p_z  h^{*-1}$ where $h(z)\in G^c$.
The action then\refmark{\GK,\KS}
reads $$S_k(g,A) =S_k(g) -S_k(hh^*) \eqn\mishwzwh$$ The Jacobian  of
the change of variables introduces a   dimension $(1,0)$ system of
anticommuting ghosts $\chi$ and $\rho$ in the adjoint representation of the
group. The quantum action thus
 takes the form   of
$$S_k(g,h,\rho,\chi) =S_k(g) -S_{k+2C_G}(hh*) -i\int d^2z Tr[\rho\bar \p
\bar\chi
+ c.c] \eqn\mishwzwh$$ where $C_G$ is the second Casimir of the adjoint
representation. The second term can be viewed as $S_{-(k+2C_G)}(h)$.
Since the Hilbert space of the model decomposes into holomorphic and
anti-holomorphic  sectors  we restrict our discussion only to  the former.
There are three sets of holomorphic $G$ transformations which leave \mishwzwh\
invariant $\delta_J g=i[\epsilon(z), g] $   $\delta_I h=i[\epsilon(z), h] $
and $\delta_{J^{(gh)}} \chi^a = i\f\epsilon^b \chi^c$ ; $\delta_{J^{(gh)}}
\rho^a
= -i\f\epsilon^b \rho^c$ with $\epsilon$ in the algebra of $G$. The
corresponding currents $J^a$, $I^a$ and ${\jg}^a=i\fcr$ satisfy the $G$
Kac-Moody algebra with the levels $k$,$-(k+2c_G)$ and $2c_G$
respectively. We define now $\jt{} {a} $
$$\jt {} {a}   =J^a +I^a +{\jg}^a= J^a +I^a +i\fcr \eqn\mishbJ$$
which obeys a Kac-Moody algebra of level
$$ k^{(tot)} = k  -(k+2c_G) + 2c_G =0. \eqn\mishbk$$
The energy-momentum  tensor  $T(z)$ is a sum of  Sugawara terms of the $J^a$
and  $I^a$ currents and the usual contribution of a $(1,0)$  ghost system,
namely\refmark{\GK,\KS}
$$T(z) = {1\over k+c_G }:J^a J^a: -
{1\over k+c_G }:I^a I^a: +\r  {} a \p \c   {} a. \eqn\mishbT$$
The corresponding  Virasoro central  charge  vanishes
$$ c^{(tot)} = {k d_G\over k+c_G}  -{(k+2c_G) d_G\over -(k+2c_G)+c_G}
-2d_G =0 \eqn\mishbk$$
This last property is a  first indication that the \G model is a TCFT.
In fact  it is easy to realize that the basic  algebraic structure of
a TCFT is obeyed by the model.
This is expressed in terms of two bosonic and two fermionic operators.
The former are  $T(z)$ and the ``ghost number current" $J^\#= \c a {}
\r {} a $. The
fermionic currents are a dimension one current which is the BRST current $\jb$
and a dimension two operator $G$. These  holomorphic symmetry  generators
are given by
$$ \eqalign{ \jb=&\c a {} [J^a + I^a + \half {\jg}^a]  \cr
G=&{1\over k +c_G}\r a {} [J^a - I^a ]  \cr}\eqn\mishbGJ$$
The TCFT algebraic structure now reads
 $$\eqalign{ T(z) =&\{ Q, G(z)\} \cr  \jb =&\{ Q ,
j^\#(z)\}\cr}
\eqn\mishbalgebra$$
where $Q=\int \jb(z)$ is the BRST charge. In addition to $T(z)$ and
$\jb(z)$, the total curent $\jt a {} $ is also BRST exact,
$$ \jt {} a (z) = \{ Q , \r {} a  \}.  \eqn\mishbjt $$

The   extraction of physical states  as elements of the BRST \co  will be
the subject of the next section. We summarize here, following ref. [\SY], the
picture
emerging   from the investigation   of the torus partition
function.  The latter is expressed as\refmark\GK
$$ {\cal Z}_{G\over G} =c \tau_2^{-r}\int du  {\cal Z}^g(\tau,u) {\cal
Z}^{hh^*}(\tau,u)  {\cal Z}^{gh}(\tau,u) \eqn\mishbZ$$
where $du$ is the measure over the flat gauge connections on the torus and $r$
is the rank of $G$. ${\cal Z}^g(\tau,u)$  is the torus partition function of
the
$G_k$ WZW model
$${\cal Z}^g(\tau,u) = (q\bar q)^{-c\over 24}\sum_{\lambda_L,\lambda_R}
\chi_{k,\lambda_L}(\tau, u )\bar\chi_{k,\lambda_R}(\tau, u )
N_{\lambda_R,\lambda_L}\eqn\mishbZg$$ where $q=e^{2i\pi\tau}$,
$\lambda_R,\lambda_L$ denote the $G_k$ highest weights,  and for
the diagonal invariant $N_{\lambda_R,\lambda_L}=\delta_{\lambda_R,\lambda_L}$
The character can be written as
 $$\chi_{k,\lambda}(\tau, u ) ={M_{k,\lambda}(\tau,u)\over
M_{0,0}(\tau,u)},\eqn \mishbchi$$
with $ M_{k,\lambda}$ defined explicitly for the $SU(2)$ case below,
${\cal Z}^{hh^*}(\tau,u) $ in eqn. \mishbZ\ is the contribution of
$h\in {G^c\over G}$ at level $k+2C_G$ or equivalently $h\in G$ at level
$-(k+2c_G)$. This was calculated in ref. [\GK]  using the iterative Gaussian
path
integration technique. The outcome is that
${\cal Z}^{hh^*}(\tau,u)  \sim |M_{0,0}(\tau,u)|^{-2}$ indicating that
${G^c\over G}$ contains just one conformal block. It is straightforward to
calculate ${\cal Z}^{gh}(\tau,u)$, the ghost contribution to the partiton
function  ${\cal Z}^{gh}(\tau,u)\sim | M_{0,0}(\tau,u)|^{4}$ . The cancelation
of the $|M_{0,0}(\tau,u)|$  factors in eqn.\mishbZ\ is similar to the
cancelation of the $\eta$ factors in the torus partition function of $c\leq 1$
models coupled to $ 2d$ gravity.\refmark\BerK In both  cases the resulting
character  is given by the numerator of the character of the ``matter" sector.
In the \G model it is  $M_{k,\lambda}$. This cancelation property leads to an
index interpratation for $M_{k,\lambda}$. For $G=SU(2)$ this amounts to
expressing $$M_{k,j}(\tau,\theta) =\sum_{l=-\infty}^{\infty}
q^{(k+2)(l+{j+\half\over (k+2)})^2}sin\{\pi\theta[(k+2)l +{j+\half}]\}
\eqn\mishbM$$
 as
  $$  M_{k,j}(\tau,\theta) = {1\over 2i} q^{(j+\half)^2\over (k+2)}e^{i\pi
\theta(j+\half)} Tr[(-)^Gq^{\hat L_0}e^{i\pi\theta \hat J^0_{(tot)}}]
\eqn\mishbMT$$ where $\theta$ is the holonomy in the $\tau_2$ direction,
$G$ is the ghost number, $\hat L_0$ is the excitation  level and
$\hat J^0_{(tot)}$ is the $J^0_{(tot)}$ eigenvalue of the excitation.
The prefactor in front of the trace was chosen to agree  with the definition of
the vacuum we  will use in section 3 where the cohomology is worked out
and eqn. \mishbMT\ is verified.
Note that $M_{k,j}(\tau,\theta)$ is obtained from the torus
$M_{k,j}(\tau,u)$\refmark\GK by restricting to just one angle.\refmark\SY This
amounts to consider the propagation along a cylinder rather then around the
torus. As long as we are interested in the spectrum it is sufficient to
consider $M_{k,j}(\tau,\theta)$.
 This index
interpretation enables us to read important information about the physical
spectrum from eqn. \mishbM. For a positive integer $k$, $2j=0,...k$. Hence
there are  $k+1$ zero ghost number primary states which correspond to the first
term in the $q$ expansion of the different $M_{k,j}$'s i.e the term
corresponding to $l=0$ with $\hat L_0={j(j+1)\over k+2}$.
 On each of these states there is a whole tower of
states correponding to the higher terms in the $q$ expansion,  we will refer
to these states as descendants.
 In ref. [\SY ]
it was argued that those states apear for all ghost number, and for a given
$j$ and a given ghost number there
is precisely one physical state.
The \G characters are orthonormal in the $du$ measure\refmark\GK
  $$\int du
M_{k,j} (\tau, u) \bar M_{k,j'}(\bar \tau, \bar u)
=\delta_{j,j'}\eqn\mishborth$$
where $j$ and $j'$ are $SU_k(2)$ multiplets. Thus eqn. \mishbZ\ yields
for $G=SU_k(2)$ ${\cal Z}_{G\over G}  =k+1$. For integer $k$  this is precisely
the number of conformal blocks.
\def\co{cohomology\ }
\def\F{{\cal F}(J,I)\ }
\def\Frs{{\cal F}_{r,s}\ }
\def\c#1#2{\chi_{#1}^{#2}}
\def\p#1#2{\phi_{#1}^{#2}}
\def\b#1{\beta_{#1}}
\def\g#1{\gamma_{#1}}
\def\r#1#2{\rho_{#1}^{#2}}
\def\jt#1#2{{J^{(tot)}}_{#1}^{#2}}
\def\y{|phys>}
\def\t{\tilde}
 \def\pa{\partial}
\section{  $A_1^{(1)}$ level $k$ BRST cohomology on the Fock space }
We now proceed to extract  the physical states of the \G\
theory for $G=A_1^{(1)}$.  We
use the  BRST procedure to quantize the system and thus  the
physical states   are in  the  \co of  $Q$, the BRST  charge , $|phys>\in
H^*(Q)$.
  Expanding the currents $J^a,\  I^a$ and the
$(1,0)$  ghost fields $\rho^a,\ \c { } a$ in modes and inserting them   into
eqn. \mishbGJ\   we obtain the  following BRST charge
$$ Q=\sum_{n=-\infty}^\infty [ g_{ab}\c n a( J_{-n}^b+I_{-n}^b) -\half
f_{abc}\sum_{m=-\infty}^\infty  :\c{-n} a\c{-m}  b  \r{n+m} c: ]\eqn\mishQ$$
where : : denotes normal ordering namely putting  modes with negative
subscripts
to the left of those with positive ones and $\r 0 a$ to the right of $\c 0 a$.
Since  both  $\jt n a $ and $L_n$ are  $Q$ exact
namely  $$ \{Q, \r n a \}  = \jt n  a    \qquad\{Q, G_n \}  =
L_n\eqn\mishQGQQR$$  it follows that
$$L_0\y =  0 \qquad \jt 0 0 \y =  0 \eqn\mishL.$$
 For non-vanishing
eigenvalues of $ L_0$  and $\jt 0 0$  it is easy to see that  $\y$ is
in the image of $Q$ which cannot be true for  a non-trivial
 $|phys>\in H^*(Q)$.

Let us now select a sub-space \F
of the space  of physical states on which   $\r 0 0
=   0$ in addition to   $\jt 0 0 = L_0 = 0$.
 On this sub-space $Q$  which may be written as

$$\eqalign{Q  &= \c 0 0 \jt 0 0  + M \r 0  0 +\hat Q \cr
M &= -\half f_{0bc}\sum_{n\neq 0}  :\c{-n} b\c{n}c :
-\half f_{0bc}  :\c{0} b\c{0}c :,\cr} \eqn\mishhatQ$$  equals
 $\hat Q$.
  We thus start by deducing  $H^*(\hat Q)$ the \co of  $\hat Q$. The states
which correspond  to the latter  are built on a vaccum  $|J,I>$  obeying the
following relations
 $$\eqalign{J_{n>0}^a|J,I>=&0  \qquad J_0^+|J,I>=0 \qquad
J_0^0|J,I>=J|J,I>\cr
I_{n>0}^a|J,I>=&0\qquad I_0^+|J,I>=0 \qquad
I_0^0|J,I>=I|J,I>\cr
\c{n>0} a|J,I>=&0\qquad\r{n\geq0} a|J,I>=0.\cr}\eqn\mishIJ$$ A convenient way
to
handle the $J^a$ and $I^a$ currents is to invoke the following
``bosonization"\refmark\Wak $$\eqalign{J_n^+ &=\b n\cr
J_n^0 &=\sum_m :\b m \g {n-m}:  +{a\over \sqrt{2}}\phi_n\cr
J_n^- &=-\sum_{k,m} :\g m \g k \b {n-m-k}:  -\sqrt{2}a\sum_m\phi_m\g
{n-m} +kn\g n \cr}\eqn\mishwak$$ where $a^2=k+2$. The fields $\beta$
and $\gamma$ form a bosonic $( 1,0)$ system with $[\g m, \b n
]=\delta_{m+n}$. The modes $\phi_n$ correspond to   the dimension one
operator $i\pa\phi$ and they  obey  $[\p m {} , \p n {} ]=
m\delta_{m+n}$.
  In the $I$ sector a similar parametrization  of the  currents in terms of
$\t{\b {} },\t{\g   {}}$ and $\t\phi$  is performed but now with $
I_n^0\leftrightarrow  -I_n^0\ , I_n^+\leftrightarrow  I_n^- $,
which is an automorphism of the Kac-Moody algebra. In the $I$ sector we take
 $  k\rightarrow -k-4 $ and therefore $a\rightarrow ia$ . It is easy to realize
that the conditions of  eqn.\mishIJ\ are obeyed only provided  $\b
0|J,I>=\t\g 0|J,I>  =0$. The normal ordering, however, is with respect
to the usual  $SL(2,R)$ invariant vacuum   $\b
0|J,I>=\t\b 0|J,I>  =0$. In terms of the new variables, $\hat Q$ takes the form
$$\eqalign {
\hat Q =&\sum_{l,m,n} \c {-n} - [\b n  - a( \p m + - \p m - )\t\g {n-m}
-: \t\g m \t\g l \t\b {n-m-l}:
-(k+4)n\t\g n ] \cr
+&2 \sum_{n\neq 0,m} \c {-n} 0[ :
\b m\g {n-m}: -:\t\b m \t\g {n-m} :+ a\p n - ]\cr
+&\sum_{k,m,n}\c {-n} + [\t\b n -a ( \p m + + \p m - )\g {n-m}
-: \g m \g k \b {n-m-k}: +kn\g n ] \cr
-&{\sum_{m,n}}'\half f_{abc} :\c{-n} a  \c {-m} b \r{m+n} c:
\cr}\eqn\mishhatq$$

where  $\p n {\pm} ={1\over \sqrt{2}}(\p n {}  \pm i\t\p n {})$ and $\sum'$
denotes  a sum over $m$ and $n$ which does not include $\c 0 0 $ and $\r 0 0 $
modes.

We now proceed following ref. [\BMP] to assign a degree  to the various
fields. The idea is  to decompose  $\hat Q$ into terms of different
degrees
  in such a way that there is a nil-potent operator that carries the
lowest degree  which is zero.

An assignment that obeys this requirement is the following
$$\eqalign{ &deg (\c {} {}) =deg (\g  {}) =deg (\t\g  {}) =deg (\p {} +) =1\cr
 & deg (\r {} {})= deg (\b {} ) =deg (\t\b {} ) =deg (\p {}
 -) = -1\cr}\eqn\mishdeg$$

The  decomposition of $\hat Q$ to different degrees now reads

$$\eqalign {\hat Q=& Q^{(0)} +Q^{(1)} +Q^{(2)} + Q^{(3)}\cr
Q^{(0)} =&\sum_n \c {-n} - \b n +2a \sum_{n\ne0} \c {-n} 0 \p n - + \sum_n\c
{-n} + \t\b n\cr
Q^{(1)} =&a\sum_{m,n} \c {-n} - \p m - \t\g {n-m} +\sum_{n\neq 0,m } 2 \c {-n}
0
( : \b m\g {n-m}: -:\t\b m\t\g {n-m} :)\cr
 &-a\sum_{m,n} \c {-n} + \p m - \g {n-m}
-\sum_{m,n}'\half f_{abc} :\c{-n} a  \c {-m} b \r{m+n} c:\cr
Q^{(2)} =&-\sum_n\c {-n} -[\sum_{k,m} : \t\g m \t\g k \t\b {n-m-k}: +  (a (
\p 0 +-\p 0 -) +(k+4)n)\t\g n ] \cr
&-\sum_n\c {-n} +[\sum_{k,m} : \g m \g k \b {n-m-k}: +  (a ( \p 0 + +\p 0 -)
-kn)\g n ]\cr
Q^{(3)}=& -a \sum_{m,n}( \c {-n} - \p m + \t\g {n-m} +\c {-n} + \p m + \g
{n-m} ).\cr}\eqn\mishdegQ  $$
{}From the fact that terms of different degree in $(\hat Q)^2$  vanish
separately  it follows\refmark\BMP that on \F   $Q^i$ obey the following
relations
 $$\eqalign{&(Q^{(0)})^2= (Q^{(3)})^2 = 0\qquad \{Q^{(0)}, Q^{(1)}\} =
\{Q^{(2)},Q^{(3)}\}=0\cr
 &{Q^{(1)}}^2+\{Q^{(0)},Q^{(2)}\}=
(Q^{(2)})^2 + \{Q^{(1)}, Q^{(3)}\} = \{Q^{(0)},Q^{(3)}\} + \{Q^{(1)},Q^{(2)}\}
=
0 \cr}\eqn\mishQQ$$
In fact $(Q^{(0)})^2=0$ holds on the entire Fock  space.
We want now to apply the results of ref. [\BMP]  which hold only
provided that there is a finite number of degrees for each ghost number.
Recall that states in  \F are annihilated by both
$$\eqalign{\L_0 = &\hat L_0 +{1\over k+2} [J(J+1) -I(I+1)]\cr
 \hat L_0=&\sum_n n [:\b {-n} \g n : +:\t\b {-n} \t\g n : +g_{ab}:\r {-n}
a\c n  b: ] +\sum_{n\neq 0 } :\p {-n} + \p n - : \cr}
\eqn\mishlzero$$
and
 $$\jt 0 0=  J+I+\sum_n  [:\b {-n} \g n : -:\t\b {-n} \t\g n :-f^0_{bc}:\c {n}
b \r {-n} c :] .\eqn\mishjtot$$
It is clear from the expression of $L_0$ that for a given $|J,I>$ the
amount of excitations and thus the degree they carry is limited. The
restriction
of vanishing $\jt 0 0 $ further limits the contribution of the zero modes
$\g 0 $ and $\t\b 0 $. This proves that on \F and in particular for
each ghost number the  degree  carried by any state is bounded from
both sides.  Hence we can proceed and use the lemmas proven in ref.
[\BMP].
The next step is to find the \co of $Q^{(0)}$  on \F .
It is not difficult  to realize that $\hat L_0$, the contribution to $L_0$ of
the
exitations,  is $Q^{(0)}$ exact
$$ \eqalign{\hat L_0 &= \{ Q^{(0)}, \hat G^{(0)}\}\cr
\hat G_0 &=-\sum_n n[\r n + \g {-n}  + \r n - \t\g n]
 +{1\over a}\sum_{n\neq 0} \r n 0 \p {-n} +\cr}\eqn\mishatG$$
The consequence of the last relation is that $\hat L_0$ annihilates the states
in the \co of  $Q^{(0)}$ on  \F and thus there are no excitations in
$H^*(Q^{0})$. Moreover, since $L_0 =0$,  states in the latter must have either
$I=J$ or $I=-(J+1)$. Let us now extract the  zero modes contributions  to the
\co . The general structure of these states is
$$|n_\gamma , n_{\t\beta}, n_+ , n_- >=(\g 0)^{n_\gamma} (\t\b
0)^{n_{\t\beta}} (\c 0 +)^{n_+} (\c 0 -)^{n_-} |I,J>\eqn\mishzero$$
 where obviously $n_+, n_-
= 0,1$ and $n_\gamma, n_{\t\beta} $ are non-negative integers.
Using the following commutation relations
$$ [Q^{(0)},\g 0  ] =-\c 0 - \qquad [Q^{(0)},\t\b 0  ] =
\{Q^{(0)},\c 0 + \} =\{Q^{(0)},\c 0 - \}=0 \eqn\mishrel$$
and the relation $ Q^{(0)}|I,J>=\c 0 +\t\b 0 |I,J>$,
one finds that the result of  operating with $Q^{(0)}$ on  \mishzero\ is
$$ Q^{(0)}|n_\gamma , n_{\t\beta}, n_+ , n_- >=
 (-1)^{n_++1}n_\gamma| n_\gamma -1, n_{\t\beta}, n_+ , n_- + 1 > +(-1)^{ n_+ }
|n_\gamma , n_{\t\beta} + 1, n_+ +1 , n_- >, \eqn\mishQzero$$
where states with $n_\pm>1$ are identified as zero.
It is now straightforward to
deduce the Kernel and the Image of  $Q^{(0)}$.
The former takes the form
$$\eqalign{KerQ^{(0)}&=\sum_{n_\gamma , n_{\t\beta}}A_{n_\gamma , n_{\t\beta}}
|n_\gamma , n_{\t\beta}, 1 , 1 >\cr
&+\sum_{  n_{\t\beta}\geq 0,n_\gamma>0 }B_{n_\gamma , n_{\t\beta}}|n_\gamma
,n_{\t\beta},1,0> + B_{0,0}|0,0,1,0>\cr
&-\sum_{  n_{\t\beta},n_\gamma \geq0}
(n_\gamma +1)B_{n_\gamma +1, n_{\t\beta} +1}
|n_\gamma , n_{\t\beta}, 0 , 1 >\cr}\eqn\mishKer$$
 and the latter has the same terms appart from
the  $| 0, 0 ,1 , 0>$ term .
 Therefore the only possible  state in the relative cohomology
of  $Q^{(0)}$ is $\c 0 + |I,J>$, and from the condition $\jt 0 0 =0$  we find a
state only provided that $I=-J-1$ and then
$$ H^{rel} (Q^{(0)}) = \{ \c 0 + |-(J+1),J> \}. \eqn\mishrelco$$
The passage from the relative \co to the absolute one is  then  given by
$$H^{abs} (Q^{(0)}) \simeq H^{rel} (Q^{(0)}) \oplus\c 0 0  H^{rel} (Q^{(0)})
\eqn\mishabsco$$
in the same way as in ref. [\BMP].
Since the derived \co  includes a single degree, $ H^* (Q^{(0)})\simeq H^*(Q)$
in a complete analogy to the Liouville theory discussed in ref. [\BMP].
We conclude that the \co of $Q$ on the full Fock space includes states of
arbitrary $J$ with a corresponding $I=-(J+1)$ and with ghost number $G= 1,2$,
one
state at each ghost number. We shift from here on the definition of the
ghost-number so that the state   $\c 0 + |I,J>$ is at $G=0$.
We should note here that since there is a symmetry between the $I$ and $J$
sectors we could also have used an ``inverse bosonization" in the $J$ sector
and the ordinary one in the $I$ sector with the same resulting cohomology.
 So far we have
analyzed the cohomology on the whole  Fock Space. Now one has to pass to the
space of irreducible representations of the $J$ sector. For that we need
certain
results related to the singular vectors of arbitrary level $k$ of the
$A_1^{(1)}$ algebra. These results are presented in the next section. The
reader who is not intereseted in the details can skip it and move directly to
section 5.

\def\ra{\rangle}
\def\la{\langle}
\def\CU{\hbox{{$\cal U$}}}
\section{representation theory of $A_1^{(1)}$}
We give in this section a short review of the representation theory
of $A_1^{(1)}$ and of its singular vectors. The results presented
in this section will be heavily used in what follows. We follow the
presentation  of Malikov, Feigin and Fuks\refmark\MFF but we use the language
of current algebra. We concentrate on these results which will be relevant for
what follows.
Let g be the $A^{(1)}_1$ algebra
defined by the commutation relations
$$\eqalign{
[J^0_n,J^{\pm}_m]&=\pm J^\pm_{n+m}\cr
[J^+_n,J^-_m]&=2 J^0_{n+m}+nk\delta_{n+m,0}\cr
[J^0_n,J^0_m]&={1\over 2}nk\delta_{n+m,0}\cr
[k,J^a_n]&=[J^\pm_n,J^\pm_m]=0\cr}\eqn\mishalgebra$$
The universal enveloping algebra $\CU(g)$ can be written
$${\CU(g)=N_-\otimes H\otimes N_+}\eqn\eDC$$
where $N_+$ is the subalgebra generated by $J^+_0$ and $J^-_1$,
$H$ is the Cartan subalgebra generated by $J^0_0$ and $k$, and $N_-$ is the
subalgebra generated by $J^-_0$ and $J^+_{-1}$.

The highest weight vector is characterized by two parameters: the spin $J$
and the central charge $k$. It is convenient, though, to work with the variable
$t=k+2$. We denote the highest weight vector by $|J,t\rangle$. The conditions
it satisfies are
$${\eqalign{
&N_+|J,t\ra =0\cr
&J^0_0|J,t\ra =J|J,t\ra\cr
&K|J,t\ra =(t-2)|J,t\ra\cr}}\eqn\eHW $$
We recall that the Sugawara construction gives us another diagonal operator
$${L_0|J,t\ra =h(J)|J,t\ra\equiv {1\over t}J(J+1)|J,t\ra }.\eqn\eLEV$$
On this highest weight $\CU(g)|J,t\ra\sim N_-|J,t\ra$. We call this module
the Verma module and denote it $V_{J,t}$ . The Verma module is an infinite
dimensional representation of $A_1^{(1)}$.
The Verma module $V_{J,t}$ is naturally graded with respect to $L_0$ (the
level) and to $J^0_0$ (the spin). The homogeneous subspaces are finite
dimensional.  Kac and Kazhdan\refmark\KK studied
the question of reducibility of these modules and gave the following conditions
for reducibility:
$V_{J,t}$ is reducible if and only if there exist two positive integers r
and s such that the value of J is at least one of the following
$${\eqalign{
2J_{r,s,+}+1&=r-(s-1)t\cr
2J_{r,s,-}+1&=-r+st\cr}}\eqn\eKK$$
or if $t=0$.
A singular vector is a state $|\chi\ra\in V_{J,t}$ such that $|\chi\ra
\neq|J,t\ra$ and such that $|\chi\ra$ is a highest weight vector.
It is clear that a
 Verma module is reducible if and only if it contains a singular vector.
An interesting question is what are the level and spin  of a singular
vector in the module $J=J_{r,s,\pm}$ and what is its explicit form.
These questions were first studied in ref. [\MFF]. It was found that
if $J=J_{r,s,+}$ for a couple $(r,s)$  then
$${|\chi_{r(s-1),r}\ra=(J^-_0)^{r+(s-1)t}(J^+_{-1})^{r+(s-2)t}
(J^-_0)^{r+(s-3)t}\dots
(J^+_{-1})^{r-(s-2)t}(J^-_0)^{r-(s-1)t}|J_{r,s,+},t\ra}\eqn\eMSV$$
The subindices of $|\chi\ra$ keep track of the level and spin of the singular
vector
$${\eqalign{ L_0|\chi_{r(s-1),r}\ra&=(h(J)+r(s-1))|\chi_{r(s-1),r}\ra\cr
J^0_0|\chi_{r(s-1),r}\ra&=(J-r)|\chi_{r(s-1),r}\ra .\cr}}
\eqn\eLVS$$
What does this expression mean algebraically and geometrically?

 Algebraically, MFF\refmark\MFF found an infinite set of positive integer
$t$ such that this expression makes sense and it can be written as
$${|\chi_{r(s-1),r}\ra=\sum_{p,q} P_{p,q}(g([J^-_0,J^+_{-1}]),t)
(J^-_0)^{rs-p}(J^+_{-1})^{r(s-1)-q}|J_{r,s,+},t\ra }\eqn\eALM$$
where $P_{p,q}$ depend polynomially on $t$ and, thus, can   be continued
analytically to the whole complex plane. Take for example $r=1,s=2$ then we use
analytically continued commutation relations
$${[(J^-_0)^x,J^+_{-1}]=-2xJ^0_{-1}(J^-_0)^{x-1}-x(x-1)J^-_{-1}(J^-_0)^
{x-2}}\eqn\eCMR$$
in order to solve
$${|\chi_{1,1}\ra=(J^-_0J^+_{-1}J^-_0-tJ^-_0J^0_{-1}-tJ^0_{-1}J^-_0
-t^2J^-_{-1}) |J_{1,2,+},t\ra}\eqn\eVOO$$
If $J=J_{r,s,-}$ then $J^-_0\leftrightarrow J^+_{-1}$ and
$|\chi\ra=|\chi_{rs,-r}\ra$.
We remark that in practice there is another way to find the singular vector
which is more efficient  and  is briefly described  in  the appendix .

In order to see the geometrical meaning of \eMSV\ we
denote $\lambda_1=2J$ and $\lambda_2=k-2J$. Using \eCMR\ we see that
formally
$${\eqalign{(J^-_0)^{\lambda_1+1}|J,t\ra &\simeq |-J-1,t\ra\cr
(J^+_{-1})^{\lambda_2+1}|J,t\ra &\simeq|k-J+1,t\ra\cr}}
\eqn\eWR$$
are singular vectors for any value of J. The geometrical meaning of
these operators is clear. In the $\lambda$ plane (see Fig. 1) they conserve
the line $\lambda_1+\lambda_2+2=t$.
\vskip5cm
\centerline{Fig. 1- The $\lambda$ plane.}
In fact the operators
$(J^-_0)^{\lambda_1+1}$ and $(J^+_{-1})^{\lambda_2+1}$ are reflections of
this line on itself by the points $(-1,t-1)$ and $(t-1,-1)$ respectively.
They are the generators of the Weyl group.
MFF\refmark\MFF
 assert that for the values of $J$ that were given by Kac and
Kazhdan\refmark\KK we get after finite number of reflections a meaningful
expression.

If $\lambda_1+1$ ($\lambda_2+1$) is positive then
$(J^-_0)^{\lambda_1+1}|J,t\ra$ $\left((J^+_{-1})^{\lambda_2+1}|J,t\ra\right)$
is a formal singular vector in the Verma module $V_{J,t}$ ($V_{J,t}$). If on
the other hand $\lambda_1+1$ ($\lambda_2+1$) is negative then
$V_{J,t}\simeq (J^-_0)^{-\lambda_1-1}|-J-1,t\ra$
$\left(
V_{J,t}\simeq (J^+_{-1})^{-\lambda_2-1}|k-J+1,t\ra\right)$
is a formal singular vector in the Verma module $V_{-J-1,t}$ ($V_{k-J+1,t}$).
The only reducible modules which can not be generated by (formal) singular
vectors are those that satisfy $\lambda_1,\lambda_2\geq -1$. These
conditions for the case $t=p/q$ read
$${0\leq rq-p(s-1)\leq p}
\eqn\eCON$$
For such $J_{r,s}$ we can calculate the inclusion diagram
\vskip5cm
\centerline{Fig.-2 J inclusion diagram}
Where
$${\eqalign{a^{p,q}_{r,s}(l)&=J^{p,q}_{r-2lp,s}={r-1\over 2}-{s-1\over 2}
{p\over q}-lp\cr
b^{p,q}_{r,s}(l)&=J^{p,q}_{-r-2lp,s}={r-1\over 2}-{s-1\over 2}
{p\over q}-lp-r\cr}}
\eqn\eSVS$$
are the spins of the singular vectors. An arrow from x to y means that
y is a singular vector of x. The solid arrows are those obtained by the
Weyl reflections. The dashed arrows inclusion was proved by other method
in ref. [\RCW] as well as the fact that this list
is exhaustive.
It is interesting to notice that there exists a kind of ``mirror symmetry''
between the representations $|J,t\ra$ and the representations
$|-J-1,-t\ra$. In fact for $J_{r,s,\pm}$ that satisfy \eCON, $|-J_{r,s,\pm}-1,
-t\ra$
are irreducible modules which are included in an infinite number of Verma
modules. The inclusion diagram (Fig. 3) is similar to that of Fig. 2, only the
direction of the arrows is inverted.
\vskip5cm
\centerline{Fig. 3-I inclusion diagram}

It is even more interesting to notice the striking similarity of this
structure to the Virasoro representations. In the Virasoro case there
exists also a ``mirror symmetry'' between $|h,c\ra$ and
$|1-h,26-c\ra$.\refmark\LZ We will see in the following chapters that this
similarity is the reason for the similarity between the spectra
of ${SL(2)\over SL(2)}$ and
that of the minimal models coupled to Liouville. We refer to this mirror
symmetry as ``duality".
Note also that for $r=p$ eqn.\eSVS\ gives $a(l)=b(l+1)$. Hence, in this case,
we
have just one set of  singular vectors and the two branches of the embedding
diagram of Figs. 2,3 degenerate into a single branch diagram.
\section {  Irreducible representation and the BRST cohomology}
The next step in the extraction of the physical states is to pass
from the \co on the Fock space to  the irreducible representations
of the level $k$  $A_1^{(1)}$  Kac- Moody algebra. In general a
representation $L$ is reducible iff $ 2L +1=r-(s-1)(k+2) $ where $
r$ and $s$ are integers with either $r,s\ge 1$ or $ r<0, s\leq
0$.\refmark\KK In the \G model with $G= A_1^{(1)}$ we therefore have
$$2J_{r,s} +1 =r-(s-1)(k+2)\qquad  2I_{r,s} +1
=r+(s-1)(k+2)\eqn\mishJIrs$$ Note that $J_{r,s} =-I_{-r,s}-1$.
Completely irreducible representations, which have infinitely many
null vectors, appear provided that $k+2={p\over q}$ for $p$ and $q$
positive integers which can be chosen  with no common divisor.  In
this case  $I_{r,s} = I_{r+p, s-q}$ and $J_{r,s} = J_{r+p, s+q}$. It
is, thus, enough to analyze the domain $1\leq s\leq q$,  and we will
choose  $1\leq r\leq p-1$. This choice corresponds to the double line embedding
diagram of Fig 2. The states corresponding to $r=p$ have a single line
embedding
diagram.
 It was found that \refmark\BF  for the  case of the double line one can
construct the irreducible representation which is contained in $\Frs$, the Fock
space built on $|J_{r,s}>$. This is achieved
 via  the  \co of an
operator $Q_J$  which acts on   $\Frs$    the union of the
Fock spaces that correspond to $J_{r+2lp,s}$ and $J_{-r+2lp,s}$ for every
integer $l$.  It turns out\refmark\BF
 that the relevant information is
encoded in $H^0(\Frs,Q_J)$ and all other levels of the cohomology vanish. The
cohomology is only in the $J$ sector and not in the $I$ sector just as there is
no use of the \co of the Liouville sector in models of $C<1$  matter coupled to
gravity\refmark\BMP. Thus, the space of physical states  of ghost
number $n$ is given by $$H^{(n)}_{rel} [ H^{(0)}(\Frs,Q_J)\times {\cal
F}_I\times {\cal F}^G,Q]\eqn\mishHa$$
where  $ {\cal F}^G$ is the ghosts' Fock space built on the  new vaccum
$|0>_G$.
Since $Q_J$ acts only in the $J$ sector we can rewrite $ H^{(n)}_{rel}$ as
$$H^{(n)}_{rel} [ H^{(0)}(\Frs\times {\cal F}_I\times {\cal
F}^G,Q_J),Q].\eqn\mishHb$$ Moreover, since $\{Q,Q_J\}=0$ one can use
theorems\refmark\BT
  about double cohomologies and write this as isomorphic to
$$H^{(n)} [ H^{(0)}_{rel}(\Frs\times {\cal F}_I\times {\cal
F}^G,Q),Q_J].\eqn\mishHc$$
The theorems\refmark\BT
 apply only provided that each \co separately is
different from zero only for one single degree. In the present case this was
shown in section 3.  In fact we have already calculated $
H^{(0)}_{rel}(\Frs\times {\cal F}_I\times {\cal F}^G,Q) $ since $\Frs$ is the
union of free Fock spaces. Hence the result is that the latter has one state if
the Fock space of $J=-I-1$ is in $\Frs$, and it is empty otherwise.
For each $J_{r,s}$ we get states at $I=-J_{r+2lp,s}-1 $ and
 $I=-J_{-r+2lp,s}-1 $ where their ghost number is equal to the corresponding
degree in the complex of ref. [\BF]
{}.
For each such $J_{r,s}$ there is an infinite set of states with
$I=I_{-r-2lp,s}\ \  G=-2l $ and $I=I_{r-2lp,s}\ \  G= 1-2l$ for every integer
$l$. An example  is the case of $k=0$ which is discussed in
detail in section 7. There $J_{r,s}=0 $ since the only possible values of $r$
and $s$ are $r=s=1$. Thus the posible states are at
$I_{-r-2lp,s}=-2l-1 $ and $I_{r-2lp,s}= -2l$  which implies that the possible
values of the ghost number  are $G=I+1$ for every integer $I$.

Though the  general derivation of the \co does not give an explicit
construction of the physical states, it is clear that they involve null states.
(Recall that originally there is only one physical state in the Fock space.)
 In our construction an irreducible
representation of the $J$ sector was  used namely null states were eliminated,
while  the $I$ sector was left as a Fock space. In the latter case, it can be
shown that half of the null states vanish identically on the Fock space.
 For instance in the $k=0$ case, for
negative ghost number one has $Q|phys> = |null>$, where by $|null>$ we mean a
null state or it's descendant. Hence, these states are in the cohomology
provided the corresponding null states vanish, which indeed is the case.
  For positive ghost
number there exist states $|\psi>$ for which  $Q|\psi>=|phys> + |null>$ and
therefore those states are in the cohomology only when  the corresponding  null
states are non-zero.
 The next step
is to compute the level $\hat L_0$ and $\hat\jt  0 0 $ for the excitations
on   all states. The results are summarized as follows

$$\eqalign{ J=J_{r,s},\ \ I= I_{-r-2lp,s},\ \qquad \qquad&
J=J_{r,s},\ \ I= I_{r-2lp,s},\  \cr
 G=-2l\qquad \qquad \qquad \qquad\ \ \ \ \ \ \ \ &G=1-2l\cr\hat L_0=l^2pq
+l(qr-sp) +lp \ \qquad\qquad& \hat L_0=l^2pq -l(qr+sp) +r(s-1)+lp\ \ \cr
 \hat\jt 0  0 =lp\qquad\qquad \qquad \qquad\ \ \ \ & \hat\jt 0   0 =lp-r
\cr}\eqn\mishlevel$$ where $\hat L_0$ is given in eqn. \mishlzero\ and
$\widehat\jt 0  0 = -I-J-1 $ is the total charge of the excitations since the
total $J_0$
 vanishes and our ghost vacuum has $J_0=1$.
Again we should note that had we used the  reversed parametrization, the
results would have been analogous to those just derived. Since the direction of
the \co in the complex of ref.[\BF]  would have been reversed, we would
have obtained  the same states but with opposite ghost numbers,
namely, $G=2l$ and $G=2l-1$ instead of  $G=-2l$ and $G=1-2l$
respectively. The partition function that is computed below is not
affected by those changes.

  The last step in the
deduction of the space of physical states is the passage to the absolute \co
which is the same as in ref. [\BMP] since, as we have just shown,
 there is a single state for each $I$ and $J$,
$$H^{(n)}_{abs} \simeq H^{(n)}_{rel} \oplus\c 0 0
H^{(n-1)}_{rel}.\eqn\mishHabs$$

Let us now examine whether we can verify   the index
interpretation of the torus partition fucntion which was
discussed in section 2.
 For integer $k$ $j=0,..,{k\over
2}$. The partition function in terms of the characters $M_{k,j}$ was given in
eqn. \mishbM. We want to check now whether it can be rewritten as a trace
over the space of the physical states.
 One has to  insert the  values of $\hat L_0$ and  $\hat J^0_{(tot)}$ of eqn.
\mishlevel\ into eqn. \mishbMT, with $\widehat J^0_{(tot)}$ and $G$ shifted
to the
values defined for  an $SL(2)$ invariant vacuum. Inserting these  values for
every
  $l$ one gets exactly the expression of eqn.\mishbMT. We can obviously
add the values of the level and the eigenvalue of $\jt {} 0$ of the $|J>$
vacuum
namely   $\hat L_0 \rightarrow \hat L_o +{J(J+1)\over k+2}$ and
$\widehat\jt {} 0
\rightarrow \widehat\jt {} 0 +J$
to  derive a simpler expression $$  M_{k,j}(\tau,\theta) = {1\over 2i}
q^{1\over
4(k+2)}e^{i\pi {\theta\over 2}} Tr[(-)^Gq^{\hat L_0}e^{i\pi\theta \widehat
J^0_{(tot)}} ].\eqn\mishbMTT$$

\section{ The Correspondence to  $c<1$ models coupled to gravity}
Let us now raise the question of whether one  can  map   the ${SL(2)\over
SL(2)}$ model to  minimal models coupled to gravity. Or differently to what
extent are the two types of TCFT equivalent. The minimal models can be
formulated either in a Liouville approach\refmark\DDK or in a world-sheet
light-cone gauge.\refmark\Pol Since   these two gauges are believed to be
equivalent\refmark\MO  it  is enough to demonstrate the correspondence of the
\G
model with one of the two. Nevertheless, we discuss here the relations of the
\G  with the two  2d   gravity pictures.

Motivated by the comparison of the partition functions of the \G model of level
$k={p\over q}-2$ and that of a $(p,q)$ model coupled to gravity, which is
discussed below, we  define now a twisted energy-momentum tensor $\t T$  as
follows:  $$T(z)\rightarrow \t T(z) = T(z) +\pa\jt {} 0 (z)\eqn\mishetT$$
Since both  $T(z)$ and $\jt {} 0 (z)$ are BRST exact so is $\t T(z) $. Hence,
 this twist does not introduce a Virasoro  anomaly. However,   both the
dimensions and the contributions of the various currents to the
central charge  are now altered. In addition it is clear that $\t  T$
is not an $A_1^{(1)}$ invariant operator.  The contributions  to $c$ of
 $A_1^{(1)}$ currents at level $k$ are shifted from ${3k\over
k+2}\rightarrow {3k\over k+2}-6k$. Hence, for the case of $k+2
={p\over q}$ one gets the following anomalies in the $J$, $I$ and
$\jg$ sectors $$ \t c_J =2 + c_{p,q}\qquad   \t c_I =2 +
(26-c_{p,q})\qquad \t c_{gh} =-30\eqn\mishetc$$ where $c_{p,q}=
1-{6(p-q)^2\over pq}$. One can rewrite the contribution to the total
central charge in the form they appear in the Liouville and the
light-cone formulations

$$ \eqalign{
c^{(tot)} &=c_{p,q}+ (26-c_{p,q} ) -26\cr
c^{(tot)} &=c_{p,q}+ ({3\kappa\over
\kappa+2} -6\kappa) -28\cr} \eqn\mishec$$
where $\kappa=-k-4$.
Obviously so far it is only a rewriting of zero and by iteself it does not
prove much. However, we want to check  whether
 one may   provide a  map between   the set of  fields of the  $A_1^{(1)} $ \G
model  and those of minimal models coupled to gravity. In particular eqn.
\mishec\  may suggest the following correspondence.
The $J$ sector contains the ``matter" degrees of freedom and a bosonic $(1,0)$
system which compensates a similar anticommuting system from the ghost sector.
 The $I$ sector is the
 $SL(2,R)$ gravity part in the approach of ref.[\Pol] or it is the Liouville
sector plus an additional commuting system of $c=2$ which again pairs with a
ghost partner.
 The $\rho, \chi$ ghosts translate into the $b,c$ system (or to
the
 ghost system of ref [\KPZ]) plus  additional  anticommuting   $(1,0)$ degrees
of freedom. This implies that the ghost vacuum transformation $|0>\rightarrow
\c 0 +|0>$ corresponds to the usual $|0>\rightarrow
c_1|0>$.
 Before discussing these kind of  realtions
let us observe another ``numerological  correspondence" between KPZ\refmark\KPZ
and the   $A_1^{(1)}$ \G model. Recall that the relation between  the
``undressed" and
``dressed" dimensions is\refmark\KPZ
   $\lambda + \Delta^{(0)} -{\lambda(1+\lambda)\over
k+2} =0$.  If we identify $J$ with $\lambda$ and take for it $J=J_{r,s}$ then
we
find that
$$  \Delta^0 = {J(1+J)\over k+2}- J = {1\over 4t}[(ts-r)^2
-(t-1)^2] =\Delta^{(0)}_{r,s}\eqn\misheDelta$$  for $t={p\over q}$.

The modified $T$ of eqn. \mishetT\ introduces the following modified
dimensions:
$$\eqalign{ (\r {} -, \c {} +) \rightarrow \ \ &(2,-1)\ \  \leftarrow(\t\b {},
\t\g {} )\cr
 (\c {} -, \r {} +) \rightarrow \ \ &(1,0)\ \  \leftarrow(\g {} ,\b
{} )\cr}\eqn\mishedim$$
The conformal dimension of $\c {} 0, \r {} 0$  and $\phi,\ \t\phi$
remains
the same.

Whereas it is obvious that $\phi$ corresponds to the Fegin-Fuks\refmark\FF free
field representation of the matter part, since their central charges are the
same,
one cannot simply assign $\t \phi$ to
 the Liouville degree of freedom. This is due to the fact that its
background charge is $\half(\sqrt{k+2} -{1\over \sqrt{k+2}})$ which leads
to a contribution to the Virasoro central charge $2-c_{p,q}$ rather than
$26-c_{p,q}$. The role of $\pa\phi_L$, where $\phi_L$ is the Liouville mode, is
played here by a combination of $\pa\t\phi$ and $\t\b {} {}\t\g {} {}$.
 Notice that  for $k=-1$ which formally correponds to $c=1$ both
$\phi$ and $\t\phi$ have a vanishing background charge.
In the reversed parametrization mentioned in section 3, $\t\phi$ would have
had the central charge of the Liouville  mode but then  the central
charge of $\phi$
would have been $c_{p,q}-24$ instead of  $c_{p,q}$. Obviously using the same
parametrization for both the $J$ and $I$ sectors implies that $\phi$ and
$\t\phi$ have the background charges of the two scalars in the $(p,q)$ models
coupled to gravity. Moreover, in this case   the systems $\b {} {}, \g
{} {} $ and  $\t\b {} {}, \t\g {} {} $ systems carry $(1,0)$ dimensions and
pair
with $\c {} -\r {} +$ and  $\c {} 0\r {} 0$.

 Next we want to compare the partition function of the $(p,q)$
model to that of \G for  $G=SL(2)$ and  $k= {p\over q}-2$. Comparing eqn.
\mishbMT\ to the
numerator of the character of the minimal model it is clear that correspondence
might be achieved only provided one shifts $\tau\rightarrow \tau-\half \theta$
or
equivalently taking $e^{i\pi\theta}=q^{-1}$. In this case  the numerator of the
character in the minimal model which is proportional to $Tr[(-1)^G q^{\hat
L_0}]$ is mapped into  $Tr[(-1)^G q^{\hat L_0 -\widehat \jt 0 0 }]$ in the \G
model. That is the origin of the twisted $T$ defined in eqn. \mishetT. We thus
need to compare the number of states at a given level and ghost number in the
minimal models with the corresponding numbers at the same ghost number and
``twisted level". From eqn. \mishlevel\  we read
 $$\eqalign{
 I= I_{-r-2lp,s}\ \ G=-2l \qquad&\hat L_0 -\widehat\jt 0 0=l^2pq
+l(qr-sp) \cr
 I= I_{r-2lp,s}\ \ G=1-2l \qquad&\hat L_0 -\widehat\jt 0 0=l^2pq
-l(qr+sp)  +rs\cr
}.\eqn\mishdL$$
In the minimal models we have states built on vacua
labeled  by the pair $r,s$ with $1\leq r\leq p-1$ and $1\leq s\leq q-1$
with $ps>qr$ which have dimension $h_{r,s} = {(qr-sp)^2-(p-q)^2\over 4pq}$.
The levels of the excitations  are $\hat L_0=\Delta-h_{r,s}$.
  For $G=2l+1$ one has $\Delta=A(l)={[(2pql+qr+sp)^2-(q-p)^2]\over 4pq}$ and
for $G=2l$ $\Delta=B(l)={[(2pql-qr+sp)^2-(q-p)^2]\over 4pq}$.
\refmark{\BerK,\LZ}
Hence,
the  the contribution of the various levels to the partition function
 are identical to those of $\hat L_0 -\widehat\jt 0 0$ in  eqn. \mishdL\ for
 the
same ghost numbers and the respective vacua satisfy $J=\sqrt{p\over 2 q}p_m$
and
$I=-\sqrt{p\over 2 q}p_L$ where $p_m$ and $p_L$ are the matter and Liouville
momenta respectively. It is thus clear that for a given $r,s$ we get the same
number of states with the
same ghost number parity in the two models and thus the two partition functions
on the torus are in fact identical.
The relation between $a_{r,s}(l)$ ($b_{r,s}(l)$)
 of eqn. \eSVS\ and the dimensions  $A_{r,s}(l)$ ($B_{r,s}(l)$) of the null
states appearing on the minimal models embedding diagram is given by
$${1\over t}a^{p,q}_{r,s}(l)(a^{p,q}_{r,s}(l)+1)-a^{p,q}_{r,s}(l)=A^{q,p}_{r,s}
(l)=A^{p,q}_{s,r}\eqn\mishabAB$$
with a simila expression relating $b_{r,s}(l)$ to  $B_{r,s}(l)$.  To obtain
the partition function of the $(p,q)$ models coupled to 2d gravity we have
restricted $r$ and $s$ as follows $1\leq s\leq q-1$ and  $1\leq r\leq p-1$. It
is interesting to note
that we could include in the sum over $r$ and $s$ which appears in the
partition function also the terms with $r=p$. Those terms arise from the
states  which have a single line as their embedding diagram. The $r=p$ terms
cancel between themselves and do not change the result for the partition
function. This cancelation follows from the observation that
$a_{r,s}(l)=b_{r,s}(l+1)$ which translates via eqn. \mishabAB\
into $A_{r,s}(l)=B_{r,s}(l+1)$. The
$(p,s)$ states appear at the boundary of the Kac table.
 Further discussion of these
states appear in the last section.

 In fact there are 4 different possible
identifications of the \G states with those of the minimal models. There are
the two possible parametrizations discussed in the previous sections, and there
is in each of them the possibility to assign $h_{r,s}$ either to $J_{r,s}$ or
$J_{p-r,q-s}$ with the appropriate ghost numbers.  The complete identification
of a minimal model coupled to gravity and its \G counterpartner requires
obviously  the identification of all physical states and all non-trivial
correlators in both theories. This question
 is under current investigation.

\def\co{cohomology\ }
\def \GW{$G-WZW$}
\def\F{{\cal F}(J,I)\ }
\def\c#1#2{\chi_{#1}^{#2}}
\def\p#1#2{\phi_{#1}^{#2}}
\def\b#1{\beta_{#1}}
\def\g#1{\gamma_{#1}}
\def\r#1#2{\rho_{#1}^{#2}}
\def\jt#1#2{{J^{(tot)}}_{#1}^{#2}}
\def\y{|phys>}
\def\t{\tilde}
 \def\pa{\partial}
\def \f{f^a_{bc}}
\def \fcr{\f \c b {} \r c {} }
\def \jg{J^{(gh)}}
\def \jb{J^{(BRST)}}
\def \p{\partial}
\section { Physical states and correlation functions in the
$ SU_{k=0}(2)$  case\refmark\Bersh}
As an explicit demonstration of the general results obtained  in the
previous section we consider  the simplest non-trivial case namely  \G
model for $SU(2)$ at level $k=0$. Here we present
the calculation of the \co
in the free field parametrization of section 3 as well as in a scheme where
both the $J$ and the $I$ sectors are parametrized according to eqn.\mishwak\
and
show how the equivalence between the two methods is utilized in the
determination of non-trivial correlation  functions. The matter sector of the
Hilbert space  consists only of the $|J=0>$ highest weight state since all
other
states are nulls . Therefore, in this casethe choice
of the bosonization in the $J$ sector does not affect the results.  The $I$
sector is associated with the $k=-4$ Kac-Moody algebra  and in addition there
is
the usual $(1,0)$ ghost system. Following the general analysis of sections 3
and
5, the operators which  furnish the \co of the $k=0$ case are found to be

$$ \eqalign{ \t V_{n-1} = &\c {} 0 \c {} + \c  {}   -  \p \c  {}   - ...
\p^{n-1} \c  {}   -  e^{  -n\t \phi}\delta(\t\gamma)\cr
\t V_{-1} = &\c {} 0 \c {} + \delta(\t\gamma)\cr
 \t V_{-n-1} = &\c {} 0 \c {} +
\r  {}   +  \p \r  {}   + ...     \p^{n-1} \r  {}   +  e^{  n\t
\phi}\delta(\t\gamma)  + corrections\cr}\eqn\mishfO$$ for $n>0$.
In the second scheme that we have used,  the same  parametrization
for both the $J$ and $I$ currents is used. It is  the one given in
eqn.\mishwak\
where now  $a = \sqrt 2$ and $-i\sqrt{2}$ for the $k=0$ and $k=-4$
sectors respectively. As mentioned earlier the passage between the two
prescriptions is via the automorphism $I^{0}\leftrightarrow -I^{0}$
and  $I^{\pm}\leftrightarrow I^{\mp}$.
The other difference is the properties of the  vacuum. The operators in section
3
were constructed with  respect to the $\t\gamma_0| 0>=0$, whereas now we use
the
$SL(2,R)$ invariant vacuum   which is annihilated by $\t\beta_0$. Formally
the relation between the two vacua can be written as $|\t\gamma_0=0>=\delta
(\t\gamma)|\t\beta_0=0>$

First it is straightforward to check that the following states are in the
cohomology in this parametrization.
$$\eqalign{ |V_0> &= |I=0>\cr
|V_n> &=  \c {0} + \c {-1} + ...\c {-(n-1)} + |I=-n>.\cr}\eqn\mishfVn$$
  Next we want to extract states with negative ghost number. The steps
in the  construction are the following. First we search for states $|V_{-n}>$
which are BRST closed up to a null state namely
$$ Q|V_{-n}> = |null>_n \eqn\mishfQV$$
where  by $|null>_n$ we denote a null state or a descendant of a null  which is
built on the state  $|I=n>$ as depicted in the embedding diagram.
In addition we pick states which carry the lowest ghost number possible
for the given  level and $\widehat\jt {} {}$,  so that they cannot be
 BRST exact.
It is easy to verify that states of the form
$|V_n>=( \r {-1} -\r {-2} -..\r
{-n}
- + corrections )|I=n>$ obey these conditions.
For instance the first state which has the explicit form
$$ |V_{-1}> = ( \r {-1} - + \r {-1} 0 I^{-}_0 -\half \r {-1} +
(I^{-}_0)^2)|I=1>\eqn\mishfV$$
leads to the null state of $r=1,s=2 $ when acted with Q namely
$$ Q|V_{-1}> = ( I_{-1}^- + I_{-1}^0I^{-}_0 -\half I_{-1}^+
(I^{-}_0)^2)|I=1>=|\chi_{1,1}> \equiv \hat\chi_{1,1}|I=1>,\eqn\mishfVone$$
{}From the general structure of singular vectors described in section 4 we can
read the eigenvalues of $|\chi_{1,1}>$,  $L_0|\chi_{1,1}>=I_0^0 |\chi_{1,1}>=0
$. The next null state $\hat\chi_{2,1}|I=2>=|\chi_{2,1}>$ corresponds to
$|V_{-2}>$ as follows $ Q|V_{-2}>= (\r {-1} - + correction)
 \\hat\chi_{2,1}|I=2>$
with the correction terms carrying $L_0=1$ and $\jt 0 0 =-1$. Notice that since
$|\chi_{2,1}>$ has $L_0=1$ and $\jt 0 0 =+1$, it is its descendant which is in
the image of $Q$. It is easy to verify that the general structure of the BRST
charge acting on the states $V_{-n}$ takes the form
$ Q|V_n>= (\r {-1}-\r {-2}  -..\r {-(n-1)} - + correction) \hat\chi_{n,1}|I=n>$

It can be  proven\refmark\BF that all the states $|\chi_{r(s-1),r}>$ with
possitive $r,s$  vanish upon using the bosonization eqn.\mishwak\
Since we are interested in the space which includes the Fock space of
the $I$ sector, it is obvious that  $|V_{-n}>\in H^*(Q)$. Thus there
is no need to perform the procedure described in ref.[\BF]  in this
sector.  Clearly restricting ourselves to the Verma modules of the
current algebra gives rise to a  different cohomology. It is
presumably a general feature of the embedding diagram that half  of
the null states vanish upon invoking the parametrization of eqn.\mishwak. This
is proven in ref. [\BF] for any $k>0$. The generalization to $k\leq 0$ seems to
follow essentially the same steps. In the present case the null states which
carry positive values of $I$, namely those which are on the right handed branch
of the dual embedding diagram (Fig 3.) vanish. The situation is reversed once
one
uses the parametrization of section 3.   Writing  the state $|I=n>$  in terms
of
the operator $e^{n\t\phi}$ acting on the $SL(2,R)$ invariant vacuum, we can now
write down the complete set of operators that span the cohomology. $$ \eqalign{
V_{n} = & \c {}   +  \p \c  {}   + ...     \p^{n-1} \c  {}   +  e^{  n\t \phi}
\cr
 V_{0} = & 1 \cr
  V_{-n} = &
\r  {}   -  \p \r  {}   - ...     \p^{n-1} \r  {}   -  e^{  n\t
\phi} + corrections \cr}\eqn\misheO$$
Note that  all these operators  carry zero dimension and thus they close
upon a ring. The multiplication operation is just the OPE, namely

$$V_m(z)V_n(\omega) = V_{m+n}(\omega) +  \{ Q, O'\}
\eqn\mishfVV$$
where $O'$ is a dimension zero and ghost number $n+m$ operator.
The $Q$ exact term does not always appear. For instance there is no
such a term for  $n, m>0$. Using OPE's we can, in principle, starting
from eqn. \mishfV\ calculate all the $V_{-n}$.

Next we proceed to the calculation of correlation functions. To
calcualte the expectation value of products of operators which are
in the BRST cohomology, one has to define the scalar product or
equivalently the notion of the conjugate to a given state. Due to
appearance of zero modes of the ghosts $\c 0 +, \c 0 0, \c 0 -$,
the zero mode of the commuting field $\t\gamma_0$ and the
background charge $-1$ of $\t\phi$, it is clear that the vacuum is
not self-conjugate, $<0|0>=0$. Instead $,<\t 0|$ the conjugate to
$|0>$ is given by
$$<\t 0 | =<0|\c 0 + \c 0 0 \c 0 - e^{-\t\phi} \delta(\t\gamma)
, \qquad \qquad   <\t 0 |0> =1 \eqn\mishfcon$$
where we have introduced formally $\delta(\t\gamma)$  to absorb the
zero mode of $\t\gamma_0$.  To compute correllators we thus define
formally another operator in the \co
$$ \t 1 (z) = \c {} + \c {} 0 \c {} - e^{-\t\phi} \delta(\gamma)=\t V_0.
\eqn\mishfone$$  and denote by $\t V_n$ the result of the  OPE of
the $\t 1$ with  $V_n$  namely
$$ \t 1 (z)  V_n(\omega) = \t V_n (\omega) + O(z-\omega).
\eqn\mishfoneV$$ We see that the  $\t V_n$ are exactly the operators
appearing in the second parametrization up to an interchange $\c {} +
\leftrightarrow \c {} -$ and $\r {} + \leftrightarrow \r {} -$.
 Using these new definitions  we can write  the
correlators of the model as follows
$$ G(z,z_1,...z_n) =<\t 0|V_n(z)\prod_{i=1}^N V_{n_i}(z_i)|0> =
 < 0|\t 1V_n(z)\prod_{i=1}^N V_{n_i}(z_i)|0>= < 0|\t V_n(z)\prod_{i=1}^N
V_{n_i}(z_i)|0> \eqn\mishfG$$
In particular it is obvious that $<\t V_{-n}|V_n> =1$ . This proves that the
states $|V_n>$ given above are not exact, since otherwise their correlations
with closed states would vanish . Notice that since all the operators in the
correlators are of zero dimension the result is a number which is independent
on
$z, z_i$,  as it is expected for a topological model.
 The ghost number of $ V_n,
\t V_n$ are $n, n+3$ and   the momentum $P_\phi$ are $ -n, -(n+1)$
respectively.
The conditions for  a non-vanishing correlator  on the sphere are
$$\sum G = 3 \qquad\qquad \sum P_{\phi} =-1\eqn\mishfcon$$ which
translate  into a single condition for $G(z,z_1,...z_n)$   namely
$n +\sum_i n_i = 0$.
Usually in TCFT there are in addition to the correlators of zero
dimension operators also those of $(1,1)$ forms. The general derivation
of the latter from the zero forms   follows the general construction of ref.
[\WD,\dvv] $$ V_n^{1,1} = \int d^2 z G_{-1}\bar  G_{-1} V_n \bar V_n
= \int d^2 z  W_n \bar W_n\eqn\mishfGV$$

The  holomorphic  part of the lowest operator takes the form
$$ W_{-1} = 2\r {} -  \r {} + I_0^-e^{\t\phi} \eqn\mishfW$$
The computation of non-trivial correlators, and the relation with the
$(2,1)$ model, namely pure topological gravity will appear in a future
publication.

\def \GW{$G-WZW$}
\def\F{{\cal F}(J,I)\ }
\def\c#1#2{\chi_{#1}^{#2}}
\def\p#1#2{\phi_{#1}^{#2}}
\def\b#1{\beta_{#1}}
\def\g#1{\gamma_{#1}}
\def\r#1#2{\rho_{#1}^{#2}}
\def\jt#1#2{{J^{(tot)}}_{#1}^{#2}}
\def\y{|phys>}
\def\t{\tilde}
 \def\pa{\partial}
\def \f{f^a_{bc}}
\def \fcr{\f \c b {} \r c {} }
\def \jg{J^{(gh)}}
\def \jb{J^{(BRST)}}
\def \p{\partial}
\section{ Summary and Discussion}
In the present work we have worked out the space of physical states of the \G
models for the case of $A_1^{(1)} $. Strictly speaking only for integer
levels we could have used $SU(2)$ gauged WZW model. For non integer $k$ one has
to adopt the $SL(2,R)$ counterpart. The extraction of these states was done in
two stages. At first  the BRST cohomology  on a Fock space   based on a ``free
fields" parametrization was derived. In order to apply a method developed in
ref. [\BMP] for the Virasoro case, we had to apply a conjugate
parametrization in the $J$ and $I$ sectors. The second stage was the
translation of  physical states from the Fock space into  the irreducible
representation of the corresponding Kac-Moody algebras. For this
procedure we implemented the structure of the singular vectors of the
$A_1^{(1)}
$ algebra for arbitrary $k$. A ``duality" between the singular vectors
associated with the Verma module of $|J,k>$ and that of $|-(J+1),-k-4>$  is
expressed in the embedding diagrams  (Fig 2,3). This is analogus to the
``duality" in the Virasoro algebra between $|\delta, c>$ and $|1-\delta,
26-c>$.

The physical states are built on the highest state   vacua $|J,I>$ where
$J=J_{r,s}$ and $I=I_{r-2lp,s}$ or $I=I_{-r-2lp,s}$ for every integer $l$ with
$G=1-2l$ and $-2l$ respectively. The $I$ values  are those of points in the
dual embbeding diagram (Fig 3). They are related to values of the points  on
the
$J$ diagram (Fig 2.) by $I=-(J+1)$. The ghost number $G$ is in fact the
``distance" between the point on the $I$ diagram and the top of this diagram.
Using the entire space of physical states we constructed explicitly the
characters of the \G theory and verified the index interpretation of  those
chracters.

Perhaps the most intriguing  and interesting observation of this work is the
intimate connection of the ${SL(2)\over SL(2)}$ models at $k={p\over q}-2$ and
the $(p,q)$ minimal models coupled to gravity. The set of primary fields
correspond to the $|J_{r,s}>$ in the $J$ sector. The fusion  rules of the
latter are discussed in the appendix and arguments are given in favor of the
conjecture\refmark\BS that  they are closed
under their O.P.E..  The primaries and their $I$ descendants at all ghost
numbers  carry conformal  dimensions which match, after twisting, those of the
minimal models. Moreover, the partition function coincides  with that of the
latter provided that a particular value of the moduli of the flat $G$
connection  i.e $u=q^{-1}$ is chosen. This amounts to shift $\hat
L_0\rightarrow
\hat L_0 -\widehat\jt 0  0 $ which would follow from twisting the
energy momentum
tensor, $T\rightarrow T+\p\jt {} 0 $. The conformal dimension of the various
fields with respect to this modified operator are shown to correspond to those
of the minimal models with an addition of two ``topological" (1,0) systems. A
complete  isomorphism between the theories would be established when  one
compares
positively the correlation function in the two theories.
A relation between correlators in the $A_1^{(1)}$ WZW model at
 level $k={p\over q}-2$ and those
of the $(p,q)$ models have been worked out within the Hamiltonian reduction
approach.\refmark\FGPP We expect similar relations to hold in the topological
version which we study in this paper.
 Alternatively, one
would like to identify  the TCFT algebra  of the two theories. These topics are
under current investigation.
Notice that there is an apprent difference in the boundaries of the Kac table
between the \G models and the $(p,q)$ models. While in the  latter there is a
double line embedding  diagram and hence states at every ghost number for
$1\leq
r\leq p-1$,  $1\leq s\leq q-1$, in our case this happens also for $s=q$.
Single-line diagrams, which  correspond to physical states only at a finite
set of ghost numbers,appear in the minimal models on both boundaries of the
Kac table namely, $r=p$ and $s=q$ whereas in the \G models they exsit only
for $r=p$. As was discussed in section 6 those differences do not affect
the equivalence of the torus partition functions of the two types of models.
Certainly the states on the boundary of the  Kac table deserve further
investigation.  An explicit   construction of the physical  states was
demonstrated for the case of $k=0$. The zero dimension operators which
correspond to the states produced can further generate  $(1,1)$ forms  which
are
in the BRST cohomology, $V^{(1,1)} =\int dz d\bar z G_{-1}\bar G_{-1}
V^{(0)}(z)
\bar V^{(0)}(\bar z).$ That
hierarchy of operators, which is typical to all TCFT,  holds for every value of
$k$. The zero forms establish a structure of a ring . It is  expected that for
the general case the zero ghost number operators form a  fusion
ring,\refmark{\SY,\Spigel,\Gepner,\NW}  i.e.  $V_{J_1}(z)
V_{J_2}(\omega)=N_{J_1,J_2,J_3} V_{J_3}(\omega)$ where the $J_i$ refer to the
matter primaries and the $N_{J_1,J_2,J_3}$ are the fusion algebra coefficients.
It would be instructive to establish this result once the operators are
explicitly constructed.

Correlators in topological 2d gravity are known to obey recursion relations.
\refmark{\Wrr,\WD}\ We expect, therefore, similar   recursion relations
 involving
correlators defined on higher genera to be derivable in the context of the \G
models. Recently, such recursion realtions were derived using Ward identities
related to the $W_\infty$ symmetry of the ground ring of $c=1$\refmark\Kleb as
well as using ``contact relations".\refmark\KMS It will be very interesting to
recast these results in the gauged WZW framework. In particular for $c<1$ the
Ward identities approach is missing in the Liouville approach. It is still not
clear what will be the fate of its \G model counterpartner. A long standing
problem of TCFTs is to write down a theory which is isomorphic to the $c=1$
model. It is not difficult to realize that the $k=-1$ model has the right
numerologics to play the role of this theory especially when the bosonization
of eqn.\mishwak\ is chosen in both sectors.
  This model as well as some others are under current
investigation.
In fact it is easy to check using eqn.\mishetc\  that the level
which corresponds
to a given $c$ of the matter sector is $k=-{1\over12}(11+c\mp
\sqrt{(c-1)(c-25)})$. Not suprisingly a  \G  model with real $k$ has a
forbidden
domain which is the familiar range of $c$,  $1<c<25$.

 This work was entirely devoted to the $A_1^{(1)}$ case. It is pretty clear
that
a great part of the results achieved here could be extended to  other
$A_{n-1}^{(1)}$
cases and maybe other algebras. Generalizing the twisting of $T$ in the form
$T\rightarrow T+\sum_i\p{\jt {} {} }^i$ where $i$ runs over the Cartan
subalgebra, one gets for $SL(N)$ in analogy to eqn.\mishetc\ the following
contribution of the ``matter" sector to $c$, $c=(N-1)[(2N^2 +2N +1)
-N(N+1)(t+{1\over t})]$ where now $t=k+N$. Here we have assumed a ghost system
of $W_N$ gravity.  This is identical to the $c$ of $W_N$\refmark\WN models. We
note that the relationship between  $SL(N)$ and $W_N$ was established also via
the Hamiltonian reduction.\refmark{\BO,\FeFr}
 Explicit discussion of these models will
appear in a future publication.
Another obvious generalization is to the case of super Lie algebras. In
particular for $G=SL(N,N-1)$ we expect to obtain the Kazama- Suzuki
${SU(N)\over SU(N-1)\times U(1)}$ matter, which has been shown\refmark\NY to
 have
super $W_N$ algebra as its chiral algebra, coupled to super $W_N$ gravity. In
this respect we also recall the work of ref. [\NW] where the embedding of \G
models into topological matter theories was obtained by twisting hermitian
symmetric $N=2$ supersymmetric coset models.
\ack{ We are indebted  to M. Bershadsky, D. Levy, N.
Marcus, Y. Oz and M. Spigelglass for many fruitful
discussions.
One of us N.S would like to thank
  the School of Physics in the
University of Tel-Aviv for its kind hospitality. He also would like to
thank
M. Bauer, P. Di Francesco, I. Kostov, M. Petropoulos and J.-B. Zuber
for discussions.}

\refout
\appendix

\centerline{Fusion rules}

We introduce a chiral primary field
$\phi^j_m(z)$ w.r.t the Virasoro algebra as well as w.r.t
$A_1^{(1)}$. It transforms as a vector under the
horizontal algebra (the zero modes algebra):
$$[J^a_0,\phi^j_m(z)]= R^a_{mn}\phi^j_n(z)$$ Following
Zamolodchikov and Fateev \REF\rFZ{ A.B. Zamolodchikov and
V.A. Fateev, Sov. J. Nucl. Phys. 43 (1986) 657.}\refend we
introduce an auxiliary parameter in order to have
$$[J^a_0,\phi^j(z,x)]= R^a(x)\phi^j(z,x)$$
where $R^a(x)$ is a differential operator.

The correspondence fields-states is given by
$\lim_{x\to 0}\lim_{z\to 0}\phi^j(z,x)|\Omega,t>=|j,t>$. Here $|\Omega,t>$ is
the
vacuum which is characterized as a highest weight state that is
annihilated by the whole horizontal algebra.

The Virasoro algebra and $A_1^{(1)}$ are related by the
Sugawara construction
$$L_n=\sum{1\over
t}(:J^0_{n-m}J^0_m:+{1\over2}:J^+_{n-m}J^-_m:+{1\over2}:J^-_{n-m}J^+_m:)$$ In
this formulation $L_{-1}$ and $J^-_0$ generate translations in z and in x
respectively. Thus, we can write
$$\phi^j(z,x)=e^{xJ^-_0+zL_{-1}}\phi^j(0,0)e^{-xJ^-_0-zL_{-1}}$$ It follows
that
$${ [J^a_n,\phi^j(z,x)]=z^n\{(-1)^{[{a+2\over 2}]}x^{a+1}{d\over dx}
+(a+1)x^aj\} \phi^j(z,x) }\eqn\eNCR$$
Let us look now at the short distance operator product expansion for these
chiral primary fields. For this aim it is more convenient to write
$$\phi^j(z,x)=z^{-h+L_0}x^{j-J^0_0}\phi^j(1,1)x^{J^0_0}z^{-L_0}$$
which is a consequence of eq. A.1 with $a=0$. Imagine that we are interested
only in the $j$ sector in the fusion of $j_0$ and $j_1$ then
$${
\eqalign{\phi^{j_0}(z,x)|j_1,t>&=\phi^{j_0}(z,x)\phi^{j_1}(0,0)|\Omega,t>\cr
&=z^{-h_0-h_1+L_0}x^{j_0+j_1-J^0_0}\phi^{j_0}(1,1)\phi^{j_1}(0,0)|\Omega,t>\cr
&\buildrel{in\ sector\ j}\over{\hbox to 30pt{\rightarrowfill}}
 \sum_{n=0}^{\infty}\sum_{m=-n}^{\infty}
z^{h-h_0-h_1+n}x^{-j+j_0+j_1+m}
\psi^{j}_{n,m}(0,0)|\Omega,t>\cr}}\eqn\eF$$
$\psi^j_{n,m}$ are fixed by the requirement that the two sides of \eF\
transform in the same manner under the Vir and $A^{(1)}_1$ algebras.
It is clear also that $\psi^j_{0,0}(0,0)$ is proportional to $\phi^j(0,0)$.
Using the Sugawara construction we can write the following recursion relation.

$${\eqalign{(nt+m(2j+1-m))\psi^j_{n,m}=
(-j+j_0+j_1+m+1)&\sum_{k+l=n\atop k\geq1}J^+_{-k}\psi^j_{l,m+1}\cr
+2(j-j_0-m)&\sum_{k+l=n\atop k\geq1}J^0_{-k}\psi^j_{l,m}\cr
+(j-j_0+j_1-m+1)&\sum_{k+l=n\atop k\geq0}J^-_{-k}\psi^j_{l,m-1}\cr}}\eqn\eRF$$

We define $A_j(n,m)=nt+m(2j+1-m)$ and introduce an order $(n_0,m_0)>(n,m)$
if $n_0>n$ or if $n_0=n$ and $m_0>m$. We see that \eRF\ is a recursion
relation since $(n,m)$ of the left hand side is bigger then all the pairs
of indices in the right hand side.
The $\psi$'s are well defined as long as $A_j(n,m)\neq 0$. In the case where
$A_j(n,m)=0$ and $m$ divides $n$ the L.H.S of \eRF\ vanishes and we
find in the R.H.S $P(j,j_0,j_1)\times ({\rm singular\ vector})$.
The equation $P(j,j_0,j_1)=0$
is then a fusion rule since $P$ should vanish for the fusion to be possible.

\def\sss{\scriptscriptstyle}
\def\la{\langle}
\def\ra{\rangle}
We use results and notations of ref.[\BS] reviewed briefly
above as a framework for discussion of the fusion rules. The
constraints on the possible operator content of a given theory are a
direct consequence of the fact that we work in an irreducible
representation. In these representations we put the singular vectors (
if they exist) to zero. $$({\rm sing.\ vector})|j_1,t\rangle =0.$$
We use the relation between states and operators  and multiply from the left
by $\phi^{j_0}(z,x)$ to get
$$\phi^{j_0}(z,x)({\rm sing.\ vector})\phi^{j_1}(0,0)|\Omega,t\rangle =0$$
we commute the Verma module operators to the left using eq.( A.1)  and we act
with the
operator which is the outcome of this manipulation on the fused $\phi^{j_0}$
and $\phi^{j_1}$. Since $\psi_{0,0}^j$ is a highest weight state different
from zero its coefficient in the expansion should vanish. This coefficient is
a polynomial in the $j$,$j_0$ and $j_1$. Its vanishing condition is the
fusion rule.
We demonstrate this procedure by an example: take $j_1=j_{1,1,+}=0$ and
a generic $j_0$. The singular vector in the Verma module $V_{j_{1,1,+}}$ is
$$|\chi_{0,1}\rangle=J^-_0|j_{1,1,+},t\rangle$$
We have then the following equality
$$\eqalign{0&=\phi^{j_0}(z,x)J^-_0\phi^{j_1}(0,0)|\Omega,t\rangle\cr
&=(J^-_0-{d\over dx})\phi^{j_0}(z,x)\phi^{j_1}(0,0)|\Omega,t\rangle\cr }$$
We change $x\to -x$, $z\to -z$ and we multiply from the left by
$e^{xJ^-_0+zL_{-1}}$ to get
$$\eqalign{0&={d\over dx}\phi^{j_1}(z,x)\phi^{j_0}(0,0)|\Omega,t\rangle\cr
&={d\over dx} \sum_{n=0}^{\infty}\sum_{m=-n}^{\infty}
z^{h-h_0-h_1+n}x^{-j+j_0+j_1+m}
\psi^{j}_{n,m}(0,0)|\Omega,t>\cr}$$
from which the constraint
$$(-j+j_0+j_1)\psi^j_{0,0}=0$$
follows. In our case $j_1=0$ and we have the fusion rule $j=j_0$.
Another example is $j_0=j_{r,s,+}$ and $j_1=j_{1,2,+}$.
The singular vector in the Verma module $V_{j_{1,2,+}}$ is
$$|\chi_{1,1}\rangle=(J^-_0J^+_{-1}J^-_0-tJ^-_0J^0_{-1}-tJ^0_{-1}J^-_0
-t^2J^-_{-1})|j_{1,2,+},t\rangle$$
The same analysis as above gives the fusion rule:
$$(j+j_0+j_1+1)(-j+j_0-j_1)(j-j_0-j_1)=0$$
from which we conclude
$$j_{r,s,+}\otimes j_{1,2,+}=j_{r,s-1,+}+j_{r,s+1,+}+j_{r,s,-}$$
for the case $s>1$. For $s=1$ we can apply this method to the singular
vector in $V_{j_{r,1}}$ to get
$$j_{r,1,+}\otimes j_{1,2,+}=j_{r,2,+}$$
It is difficult to give the general fusion rules since we don't have an
explicit formula for the singular vectors. Nevertheless we have a conjecture
for a special class of modules\refmark\BS. For $j_1=-{n\over 2}t$ we conjecture
the condition
$$\prod_{l=-n+2\atop (-1)^l=(-1)^n}^n(lj_1+n(j+j_0+1))
\prod_{l=-n\atop (-1)^l=(-1)^n}^n(lj_1+n(j-j_0))=0$$
from which we conclude the fusion rule
$${j_{r,s,+}\otimes j_{1,n+1,+}=
\sum_{l=-n}^n{1\over 2}((-1)^l+(-1)^n)j_{r,s+l,+}
+\sum_{l=-n+2}^n{1\over 2}((-1)^l+(-1)^n)j_{r,s+l-1,-}}.\eqn\eFR$$
As we mentioned above this constraint is due to the singular vector in $j_1$.
The singular vector in $j_0$ may give further restrictions. For example it is
easy to show that if $j_0=j_{r,1,+}$ then
$$j_{r,1,+}\otimes j_{1,n,+}=j_{r,n,+}$$
 Using this result and the associativity of the OPE we finally obtain for
rational $t=p/q$ such that $j_{r,s}\equiv j_{r,s,+}=j_{p-r,q-s+1,-}$
$$j_{r,s}\otimes j_{r^\prime,s^\prime}=
\sum_{l=-s+1\atop (-1)^l=(-1)^{s-1}}^{s-1}\sum_{i=0}^{r-1}
j_{r+r^\prime-2i-1,s^\prime+l}+
\sum_{l=-s+3\atop (-1)^l=(-1)^{s-1}}^{s-1}\sum_{i=0}^{r-1}
j_{r-r^\prime-2i-1,-s^\prime-l+2}$$

Let us further remark that these fusion rules can be related to the Virasoro
fusion rules for the minimal models through the hamiltonian reduction
procedure.
Recall that in the hamiltonian reduction scheme $j$ and $h$ are related by
$$h={1\over t}j(j+1)-j$$
for $h_{r,s}$ and $t$ rational we find that both $j_{r,s,+}$ and $j_{r,s+1,-}$
are solutions. If we apply hamiltonian reduction to \eFR\ (that is, we replace
each $j$ by the appropriate $h$) we get the minimal model fusion rule.
This gives us also a clue on the way our operator should be related to the
models of minimal matter coupled to gravity. Since
$$\la \phi^{j_0}(z,x)\phi^{j_1}(0,0)\ra ={\delta_{j_0,j_1}x^{j}\over
z^{j(j+1)/t}}
\buildrel{x\to z}\over{\hbox to 30pt{\rightarrowfill}}
{\delta_{h_0,h_1}\over z^h}$$
We expect that
$$\lim_{x\to z}\phi^{j_{r,s}}(z,x)e^{\alpha\varphi}\equiv
\phi_{r,s}(z)e^{\alpha\varphi_{\sss L}}$$
in the sense of insertions in correlation functions.

\bye
\bye